\documentclass{PoS}
\usepackage{amssymb,amsmath,bm,bbm}
\usepackage{epsf,epsfig}
\usepackage{afterpage}
\usepackage{longtable}
\usepackage{cite}
\usepackage{latexsym, mathrsfs}
\usepackage{graphics}
\usepackage{url}
\usepackage{paralist}
\usepackage{bbold}
\newcommand{\be}{\begin{equation}}
\newcommand{\ee}{\end{equation}}
\newcommand{\bea}{\begin{eqnarray}}
\newcommand{\eea}{\end{eqnarray}}
\newcommand{\bec}{\begin{center}}
\newcommand{\eec}{\end{center}}

\newcommand{\vcb}{|V_{cb}|}
\newcommand{\vub}{|V_{ub}|}

\title{Theory overview of tree-level $B$ decays}

\ShortTitle{Theory overview of tree-level $B$ decays}

\author{\speaker{Fulvia De Fazio}
\\
        Istituto Nazionale di Fisica Nucleare - Sezione di Bari\\
        E-mail: \email{fulvia.defazio@ba.infn.it}}

\abstract{I describe the theoretical progress in the study of semileptonic tree-level B decays, and its interplay with recent experimental results. In particular, I focus on two anomalies: the ratios 
$R(D^{(*)})=\displaystyle\frac{{\cal B}(B \to D^{(*)} \tau \bar \nu_\tau)}{{\cal B}(B \to D^{(*)} \ell \bar \nu_\ell)}$
and the inclusive versus exclusive determination of $|V_{cb}|$.
I review a few explanations proposed for such anomalies, and discuss   tests to shed light on their origin.}

\FullConference{EPS-HEP 2017, European Physical Society conference on High Energy Physics\\
		5-12 July 2017\\
		Venice, Italy}

\begin{document}

\section{Introduction}
At present, no new particles have been observed  at the LHC, suggesting that maybe the new Physics (NP) scale is  not  directly accessible, yet.
However, a number of anomalies have appeared in  indirect searches, analyzing processes where new particles might contribute as virtual states.
From this point of view, the most promising channels are the loop-induced ones,   suppressed in the Standard Model (SM) and hence more sensitive to NP. However,  quite unexpectedly,  also tree-level processes manifest some anomalies.
Focusing on $B$ and $B_s$ decays,  observables deviating from SM predictions  in loop-induced modes  are the  angular distributions   $P_5^\prime$  in   $B \to K^* \mu^+ \mu^-$ \cite{Aaij:2015oid} and $B_s \to \phi \mu^+ \mu^-$ \cite{Aaij:2013aln}, and the ratios $R_{K^{(*)}}=\displaystyle\frac{{\cal B}(B \to K^{(*)} \mu^+ \mu^-)}{{\cal B}(B \to K^{(*)} e^+ e^-)}|_{q^2 \in [1,\,6] {\rm GeV^2/c^4}}  $ \cite{Aaij:2014ora}.
In tree-level decays, deviations have emerged in the ratios $R(D^{(*)})=\displaystyle\frac{{\cal B}(B \to D^{(*)} \tau \bar \nu_\tau)}{{\cal B}(B \to D^{(*)} \ell \bar \nu_\ell)}$ \cite{Lees:2012xj,Huschle:2015rga,Aaij:2015yra}. 
The results seem to point to lepton flavour universality (LFU) violation: 
 $R(D^{(*)})$ reveals a difference in semitauonic $B$ decays with respect to  $\mu$ and $e$ modes, and $R_{K^{(*)}}$ cast a shadow on $\mu$-$e$ universality.
This is in conflict with  the SM, where the couplings of the charged leptons  to the gauge bosons are
the same,  and LFU breaking
%difference in these interactions is due to the different value of the lepton masses.
 only arises from the Yukawa interaction.

Besides these recent experimental results, there are  long-standing puzzles in flavour physics, namely the tension  between inclusive and exclusive determinations of   $|V_{cb}|$ and $|V_{ub}|$ in the CKM matrix. It is tempting to investigate if there is a  relation with the above mentioned anomalies.
In the following, I  shall  discuss the deviations related to the semileptonic  $b \to c \ell \bar \nu_\ell$ transition. In particular, in  Section 2 I shall  focus on   $R(D^{(*)})$, briefly reviewing the experimental situation and  a few  attempts to address the anomaly. The proposal in \cite{Biancofiore:2013ki} will be discussed in some detail,  mainly because of the possibility of adopting the same framework to address also the issue of $|V_{cb}|$ determinations,  the subject of  Section 3.
Conclusions will be presented in the last Section.

\section{Ratios $R(D^{(*)})$}
Averaging the measurements of BaBar \cite{Lees:2012xj}, Belle \cite{Huschle:2015rga} and LHCb \cite{Aaij:2015yra} Collaborations,
HFAG quotes \cite{Amhis:2016xyh}:
\be
 R(D)=0.407 \pm 0.039\pm0.024  \,\,\,\, , \hskip 1 cm R(D^*)=0.304\pm 0.013\pm 0.07 \,\,\,\, , 
 \label{hfag}
 \ee
which deviate from  the SM predictions \cite{Fajfer:2012vx}:
$R(D)=0.296 \pm 0.016\,, R(D^*)=0.252\pm 0.003$ at a global 3.9$\sigma$. Moreover, a   very recent  result of Belle Collaboration, not included in (\ref{hfag}), reads   $R(D^*)=0.270 \pm 0.035({\rm stat})\pm^{0.028}_{0.025}({\rm syst})$ \cite{Hirose:2017dxl}.
%%%%%%%%

The enhancement of the tauonic mode might be explained in two-Higgs doublet models (2HDM)  introducing a new doublet of scalars with couplings to fermions proportional to their masses, as for the SM Higgs. However, BaBar   showed that it is not possible to simultaneously reproduce  $R(D)$ and $R(D^*)$ in  the simplest version of 2HDM \cite{Lees:2012xj}, hence the necessity of a
different explanation \cite{varie}.
  Efforts have been devoted  to identify  models in which the $R_{K^{(*)}}$ anomaly can be also addressed, invoking e.g. the existence of
 new  mediators  \cite{Greljo:2015mma}  or of new particles, namely leptoquarks \cite{Bhattacharya:2014wla}.
In most of such models  NP mainly couples to the third generation of fermions. Therefore, direct searches of resonances decaying to $\tau^+ \tau^-$ at LHC put stringent limits on the masses and couplings of the new mediators  \cite{Faroughy:2016osc}.
Other constraints emerge from flavour physics, in particular from meson-antimeson  mixing \cite{Buttazzo:2017ixm}. In some  NP scenarios such constraints can be safely evaded, for example in models with vector colored leptoquarks \cite{Fajfer:2015ycq}. 

A different, model-independent approach consists in considering all possible structures that can modify the SM effective Hamiltonian describing  a given process, without an {\it a priori} assumption on the NP identity, and to single out  the most sensitive observables to discriminate among the various models.
  In modes with $\tau$ leptons one can exploit observables  sensitive to the lepton mass,   unaccessible in the case of muons or electrons  \cite{Sakaki:2012ft}.
The study in  \cite{Biancofiore:2013ki} belongs to this research stream.  A tensor operator  is introduced in   the SM effective  Hamiltonian for $b \to c$ semileptonic transitions:
 \be
H_{eff}= {G_F \over \sqrt{2}}V_{cb} \left[ {\bar c} \gamma_\mu (1-\gamma_5) b \, {\bar \ell} \gamma^\mu (1-\gamma_5) { \nu}_\ell + \epsilon_T^\ell \, {\bar c} \sigma_{\mu \nu} (1-\gamma_5) b \, {\bar \ell} \sigma^{\mu \nu} (1-\gamma_5) { \nu}_\ell \right] \,\,\, , \label{heff}
\end{equation}
which affects  the ratios $R(D^{(*)})$. The new coupling
  $\epsilon_T^\ell$  is  assumed to be different for $\ell=\tau$ and $\ell=e,\,\mu$, namely 
  $\epsilon_T^{e,\mu}\simeq 0$ and  $\epsilon_T\equiv \epsilon_T^\tau$ \cite{Biancofiore:2013ki} . 
  Writing $\epsilon_T=|a_T| e^{i \theta}+\epsilon_{T0} $, 
 the data in   \cite{Lees:2012xj} on  $R(D^{(*)})$ provide the constraints  $Re[\epsilon_{T0}]=0.17 \,, Im[\epsilon_{T0}]=0$,
$|a_T| \in [0.24,\,0.27]$, $\theta \in[2.6,\,3.7]\, {\rm rad} $.
 
 Predictions for other observables can be obtained varying  $\epsilon_T$  in this range,  and allow us to discriminate the model from SM and from other NP scenarios.
One  observable is  the forward-backward  asymmetry:
${\cal A}_{FB}(q^2)=\displaystyle{\left[
\int_0^1 \,d \cos \theta_\ell \,\frac{d \Gamma}{dq^2 d \cos \theta_\ell} -\int_{-1}^0 \,d \cos \theta_\ell \, \frac{d \Gamma}{dq^2 d \cos \theta_\ell}\right]/\frac{d \Gamma}{dq^2} }$, with $q^2$ the lepton pair squared invariant mass and
 $\theta_\ell$  the angle between the  $\tau$  and $D^{(*)}$ directions  in  the lepton pair  rest frame.
\begin{figure}[!t]
 \centering
\includegraphics[width = 0.325\textwidth]{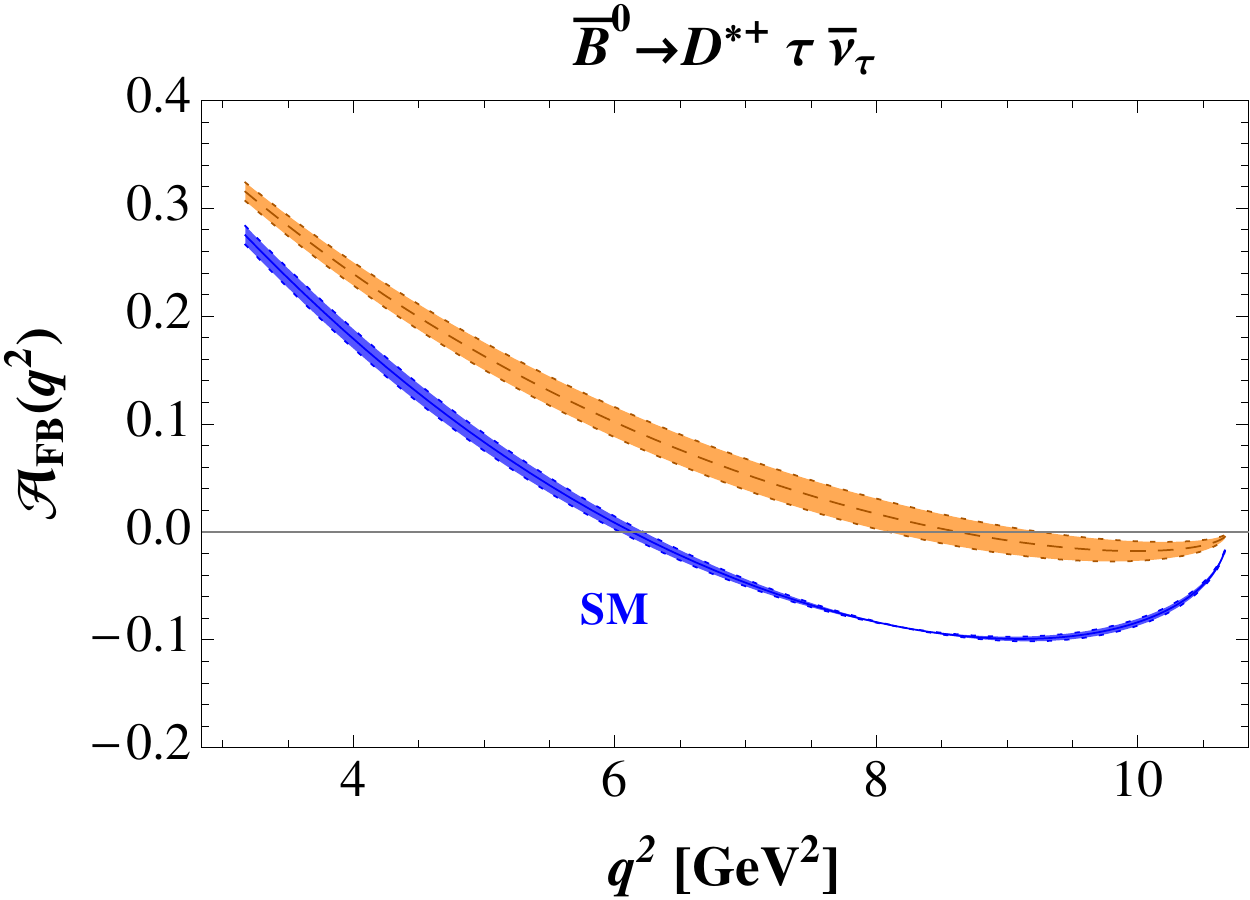}
%\\
%\hspace*{.1cm}
\includegraphics[width = 0.325\textwidth]{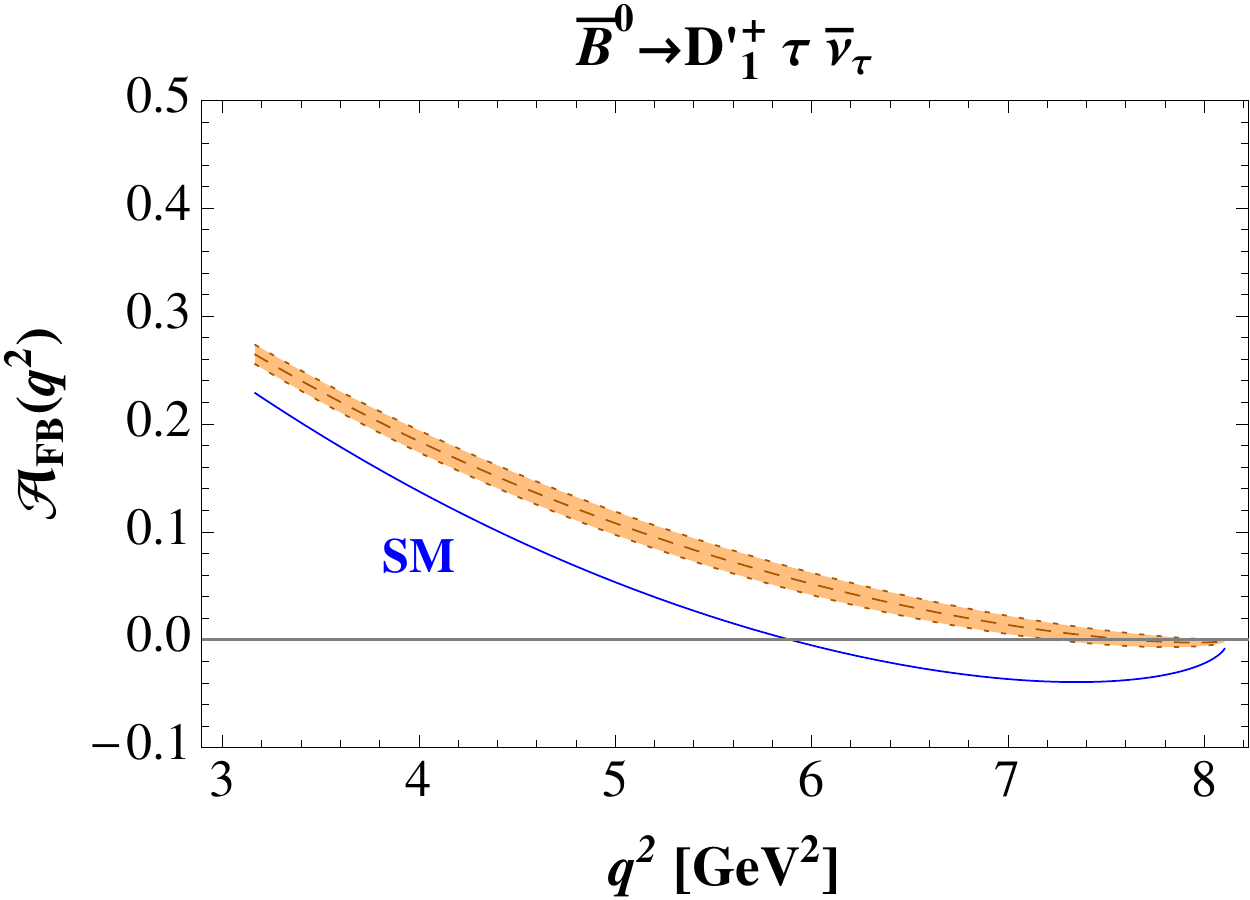}
%\hspace*{.1cm}
\includegraphics[width = 0.325\textwidth]{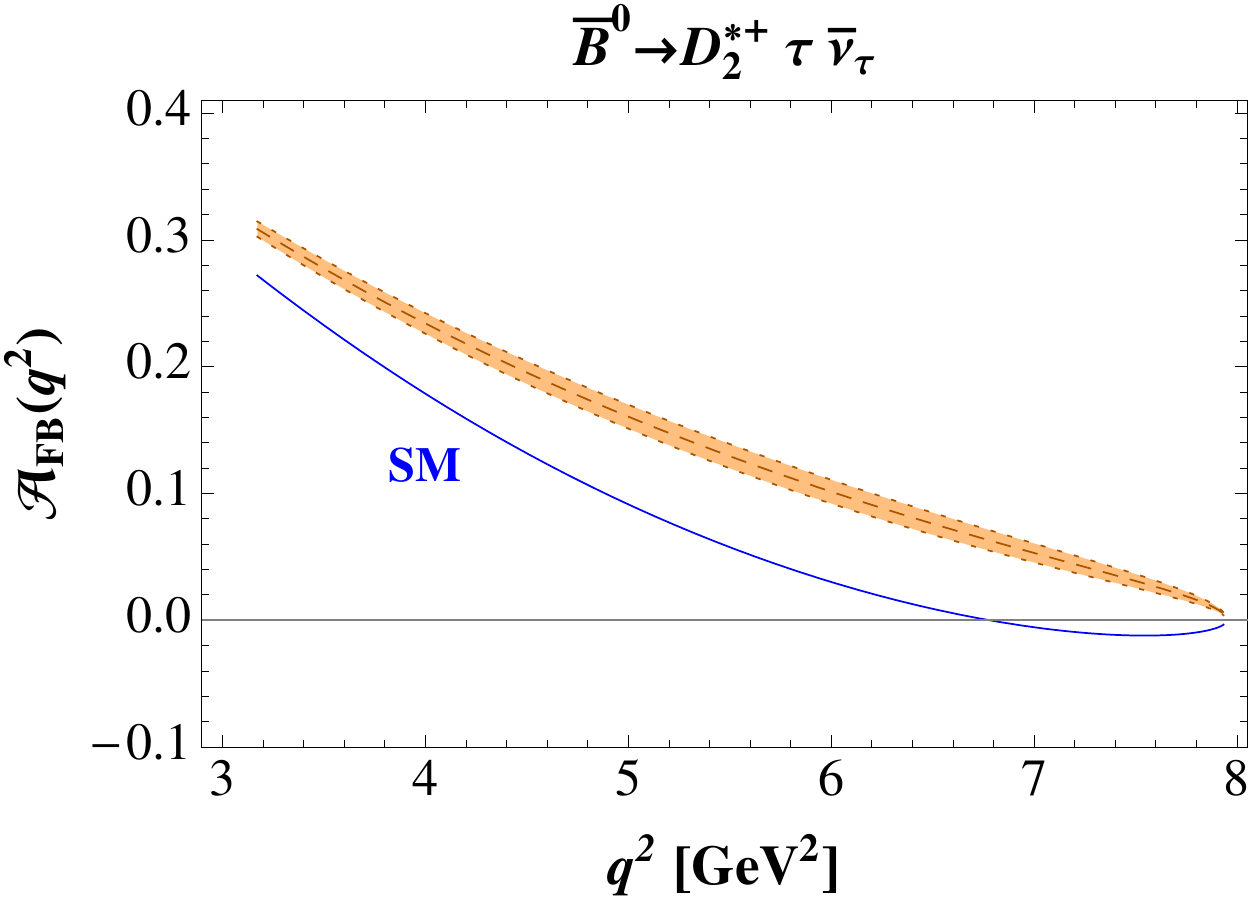}
\caption{ Forward-backward  asymmetry ${\cal A}_{FB}(q^2)$ for   $B \to D^* \tau {\bar \nu}_\tau$ (left panel),   $B \to D^\prime_1 \tau {\bar \nu}_\tau$ (middle)
 and   $B \to D^*_2 \tau {\bar \nu}_\tau$  (right panel).  In each panel the lower curve is the SM prediction,  the upper band the  NP expectation.
}\label{fig:afb}
\end{figure}
The result in  Fig.~\ref{fig:afb} (left plot), which refers to the $D^*$ mode,  shows that 
the SM prediction is systematically below the NP result, and has a zero  at  $q^2\simeq 6.15$ GeV$^2$  shifted towards larger values   in the NP model, $q^2 \in[8.1, 9.3]$ GeV$^2$.

\begin{figure}[!t]
 \centering
\includegraphics[width = 0.27\textwidth]{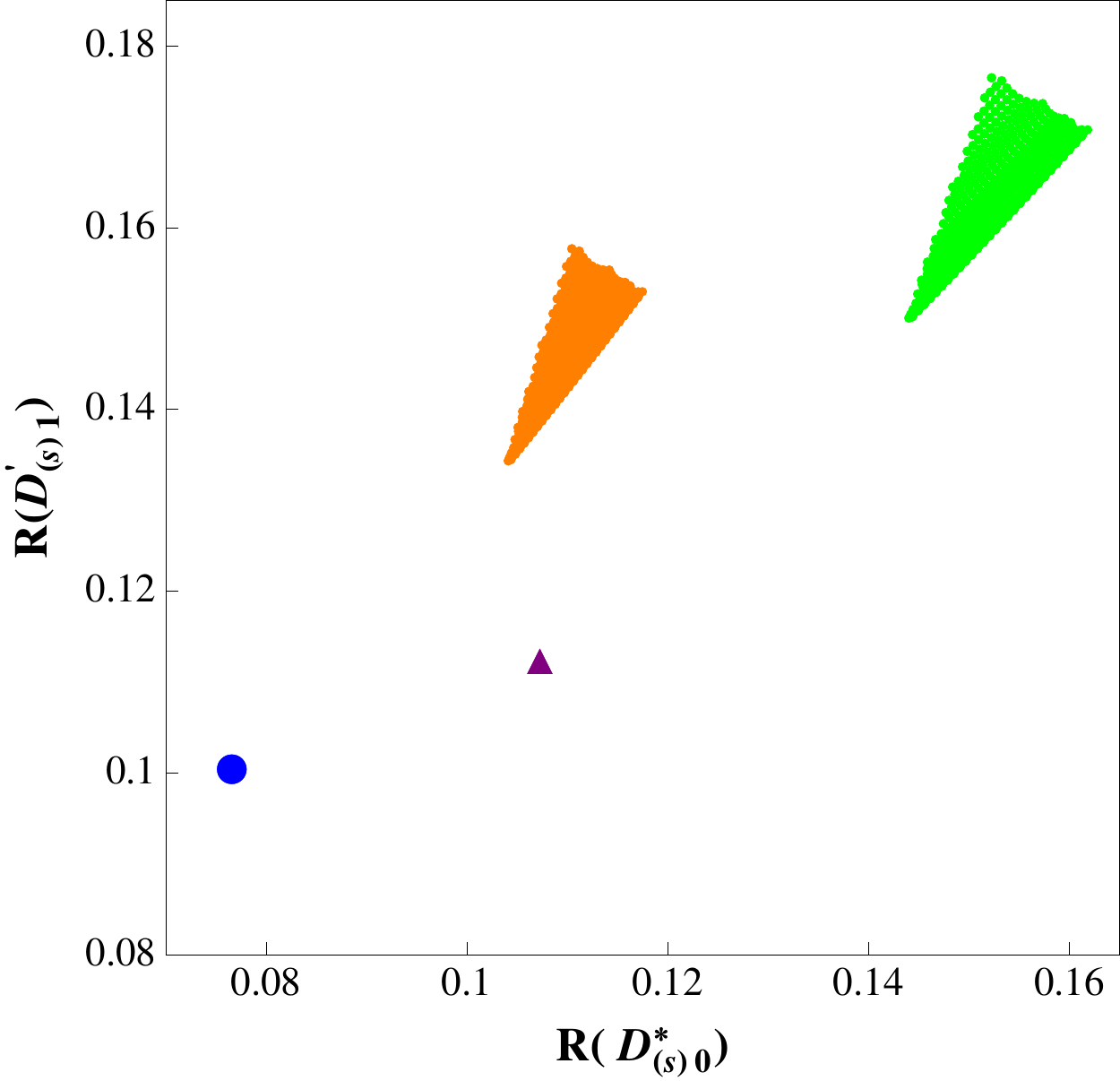}\hspace*{1.5cm}
\includegraphics[width = 0.27\textwidth]{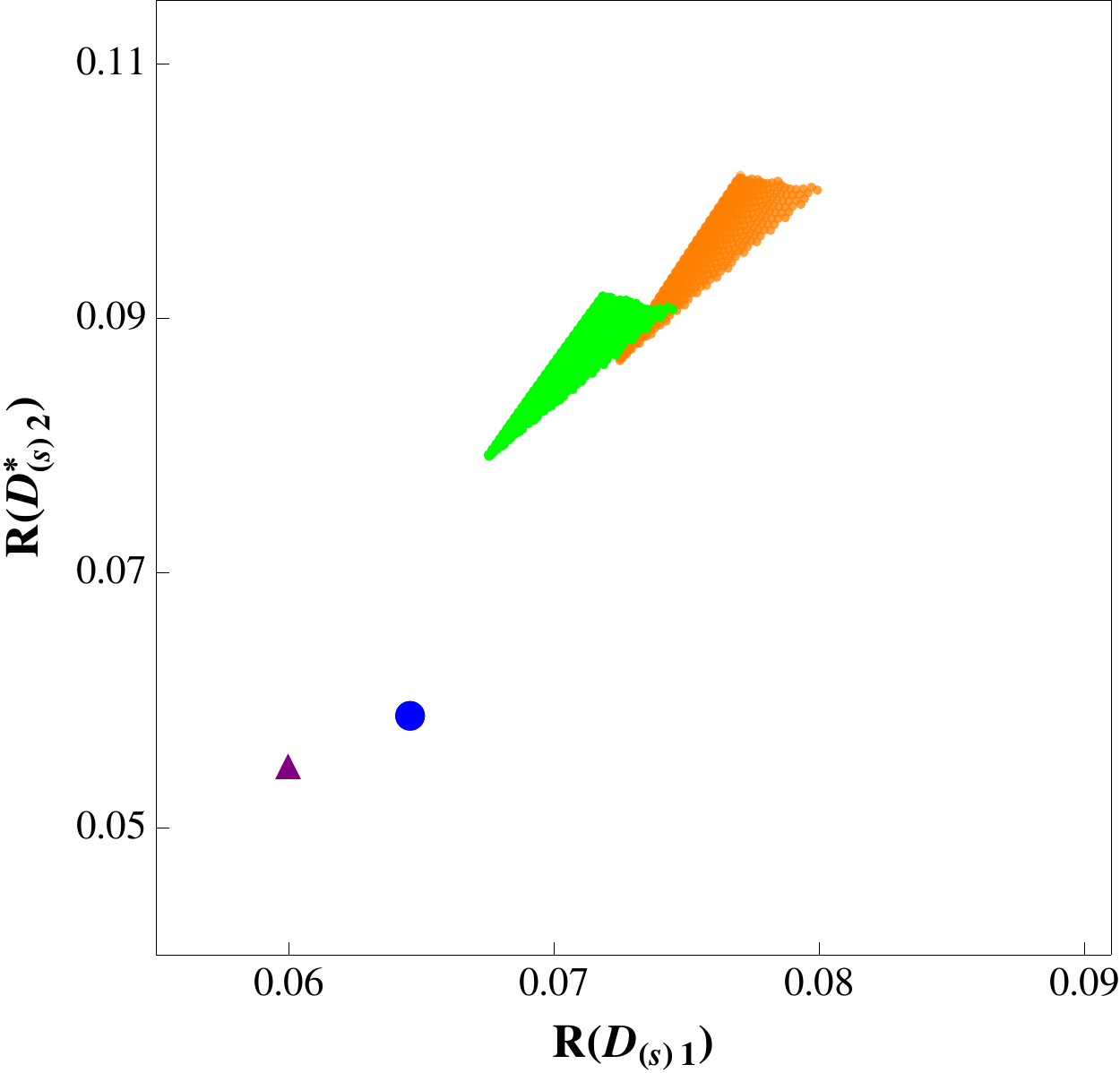}
\caption{Correlation between   ${\cal R}(D^{*}_{(s)0})$, ${\cal R}(D^\prime_{(s)1})$ (left)  and  ${\cal R}(D_{(s)1})$, ${\cal R}(D^*_{(s)2})$ (right). Orange, dark (green, light) regions refer to mesons without (with) strangeness.  The  dots  (triangles) are the SM results for mesons  without (with) strangeness.
}\label{fig:r12-r32}
\end{figure}

The tensor operator  also affects  semileptonic $B_{(s)}$ decays to  excited
charmed mesons.  The lightest  ones of such charmed states, the orbital excitations,  are generically indicated as  $D_{(s)}^{**}$ and can be classified in doublets:
 $(D^*_{(s)0},\,D_{(s)1}^\prime)$ with spin-parity $J^P=(0^+,1^+)$, and    $(D_{(s)1},\,D_{(s)2}^*)$ with $J^P=(1^+,2^+)$. All these states with and without strangeness  have been observed, as discussed in \cite{Colangelo:2012xi}.
 In the heavy quark limit,   the $B  \to D^{**}$ transitions are described in terms of two universal form factors, $\tau_{1/2}(w)$, $\tau_{3/2}(w)$.
 Using their QCD sum rule  determination   \cite{Colangelo:1992kc}
%obtained using the  QCD sum rule method  \cite{Colangelo:2000dp},
several observables can be computed, in particular  the forward-backward asymmetries. 
Fig.~\ref{fig:afb} (middle and right panels) shows the asymmetries  for   $D_1^\prime$ and
$D_2^*$. The  zero of ${\cal A}_{FB}(q^2)$, present in both cases  in SM,  is shifted in  $B \to D^\prime_1 \tau {\bar \nu}_\tau$\,  to larger values of $q^2$, and  disappears  in $B \to D^*_2 \tau {\bar \nu}_\tau$.
 The ratios 
${\cal R}(D^{**}_{(s)})=\displaystyle{\frac{{\cal B}(B_{(s)} \to D^{**}_{(s)} \tau \,{\bar \nu}_\tau)}{{\cal B}(B_{(s)} \to D^{**}_{(s)} \ell \,{\bar \nu}_\ell)}}$
can also  be computed.
 The correlations between them are shown in  Fig.~\ref{fig:r12-r32} together with
the  SM predictions. The  effect of the new tensor operator is an enhancement of  the  ratios,   correlated for the two states in  the same spin doublet.
%
%%%%%%%%%%%%%%%%%%%%%%%%%%%%%%%%%
\section{ Determination of $|V_{cb}|$ from inclusive and exclusive semileptonic $B$ decays}
The CKM  elements $V_{ub}$ and $V_{cb}$ play an important role in the SM  description of CP violation in the quark sector.  However,  the determinations  of  $\vub$ and $\vcb$ from  exclusive  and inclusive  semileptonic $B$ decays are only marginally compatible. For $\vcb$ the Particle Data Group  quotes  \cite{Patrignani:2016xqp}
\be
\vcb_{excl}=(39.2\pm0.7)\times 10^{-3}   \,\,\, , \hskip 1cm
\vcb_{incl} =(42.2\pm0.8)\times 10^{-3} . \label{vcbpdg}
\ee
The tension in (\ref{vcbpdg}) represents a long-standing puzzle in flavour physics.   Within SM  a solution has been recently proposed
 \cite{Bigi:2017njr}, using  the  Belle Collaboration data for the fully differential  $B \to D^* \ell {\bar \nu}_\ell$  decay distribution \cite{Abdesselam:2017kjf}  and scutinizing two different parameterizations of the form factors in the theoretical expression: the Caprini, Lellouch, Neubert (CLN) \cite{Caprini:1997mu}, usually adopted in the experimental analyses, and the one proposed by Boyd, Grinstein and Lebed (BGL) \cite{Boyd:1997kz}.
A similar comparison has been    performed  for  $B \to D \ell {\bar \nu}_\ell$ \cite{Bigi:2016mdz}.
While both form factor parametrizations are based on unitarity and analiticity,  the CLN parametrization relies on  HQET relations,  BGL includes single particle ($B_c^{*}$) contributions. 
It is observed that  both parametrizations have the same accuracy at large recoil in reproducing the data, but BGL better follows the low-recoil results, with  a $\vcb$ determination  closer to the inclusive one.
However, the conclusion about the critical role of the form factor parametrization in the $\vcb$ anomaly  needs to be validated using different data sets from other  experiments.
The necessity of improving HQET relations has  been stressed, since the uncertainty in the exclusive determination of $\vcb$ using the CLN parametrization could be underestimated without the  inclusion of the errors on subleading  $\Lambda_{QCD}/m_Q$ terms \cite{Bernlochner:2017jka}. 

An intriguing scenario could arise if a common NP solution were found to the $\vcb$ puzzle and the $R(D^{(*)})$ anomalies.
The possibility that NP  effects could be responsible of the tension in the $\vcb$ determinations  has been  already considered  \cite{Faller:2011nj,Crivellin:2014zpa,Bordone:2016tex}, with a negative conclusion.  In particular, the argument in \cite{Crivellin:2014zpa} is that 
a new scalar or a new tensor  operator in the effective $b \to c$ Hamiltonian, for   massless leptons, induces the same effect in both exclusive and inclusive semileptonic modes at zero recoil, producing  the same changes  in $\vcb$.  On the other hand,   a new vector or axial-vector operator  would result in modified $W$ couplings and,   due to  the  SM $SU(2)$  symmetry, in  $Z$ couplings to fermions modified at a level experimentally excluded \cite{Crivellin:2014zpa}.  Notice,
however,   that the inclusive  determination of $\vcb$  relies on 
the comparison between the  theoretical and experimental  full width \cite{Gambino:2015ima}, while
exclusive modes are analyzed close to  zero recoil  \cite{Patrignani:2016xqp}.  In \cite{Colangelo:2016ymy} it has been shown, considering in particular  the case of a  new tensor operator in the weak Hamiltonian,  that
 the lepton mass is important and, for muons,  it
can produce a sizable interference  between the SM and the NP contribution, with  a different impact on the exclusive  and  the inclusive $B$ decays,   hence on  $\vcb$. This jeopardizes  the argument in \cite{Crivellin:2014zpa}.
The NP effect is different in the inclusive and exclusive modes also in  the  electron modes, where the SM and NP interference  is negligible.

\begin{figure}[t!]
\vspace*{-0.2cm} \centering
\includegraphics[width = 0.35\textwidth]{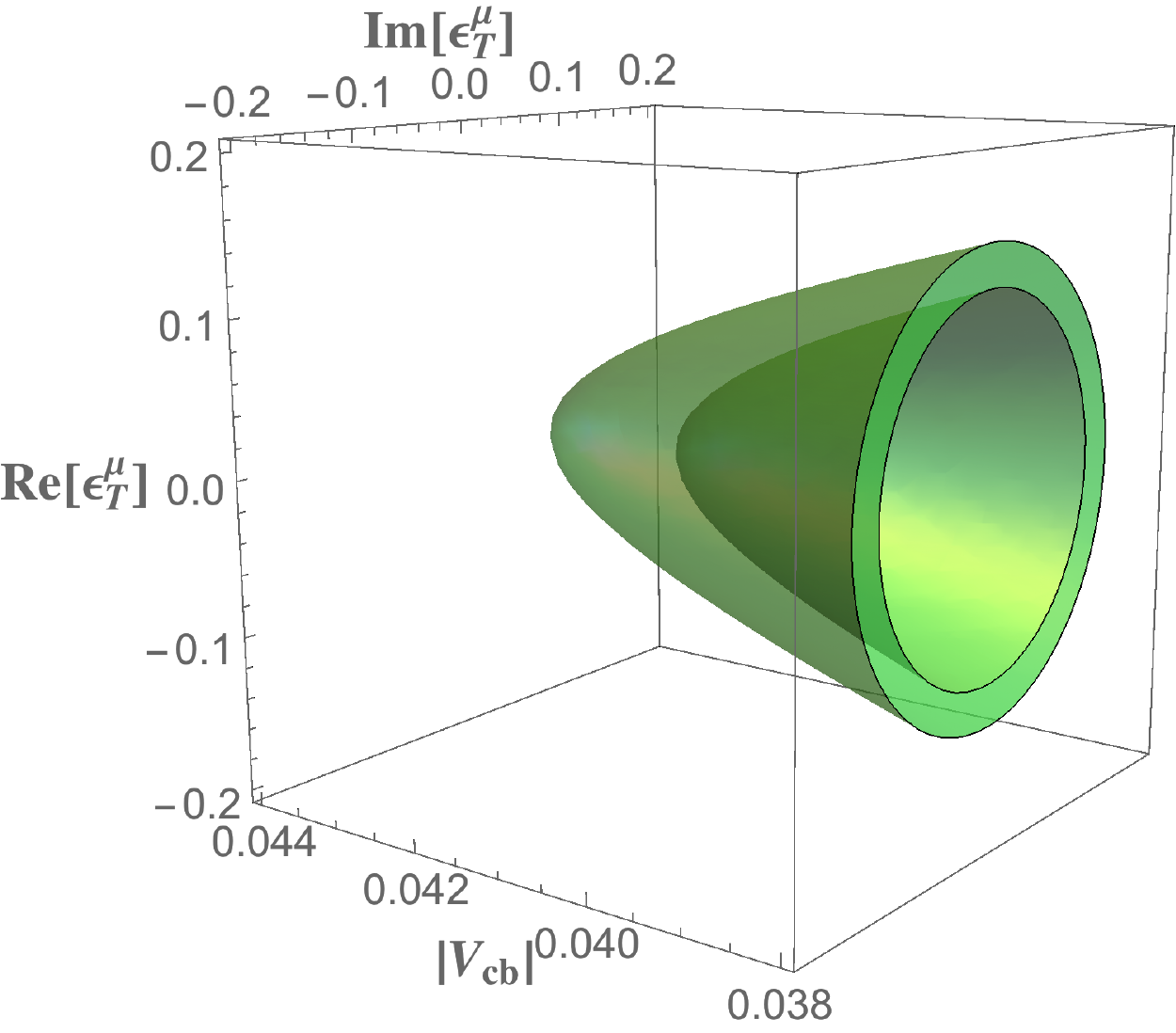} \hskip 1.5cm
 \includegraphics[width = 0.35\textwidth]{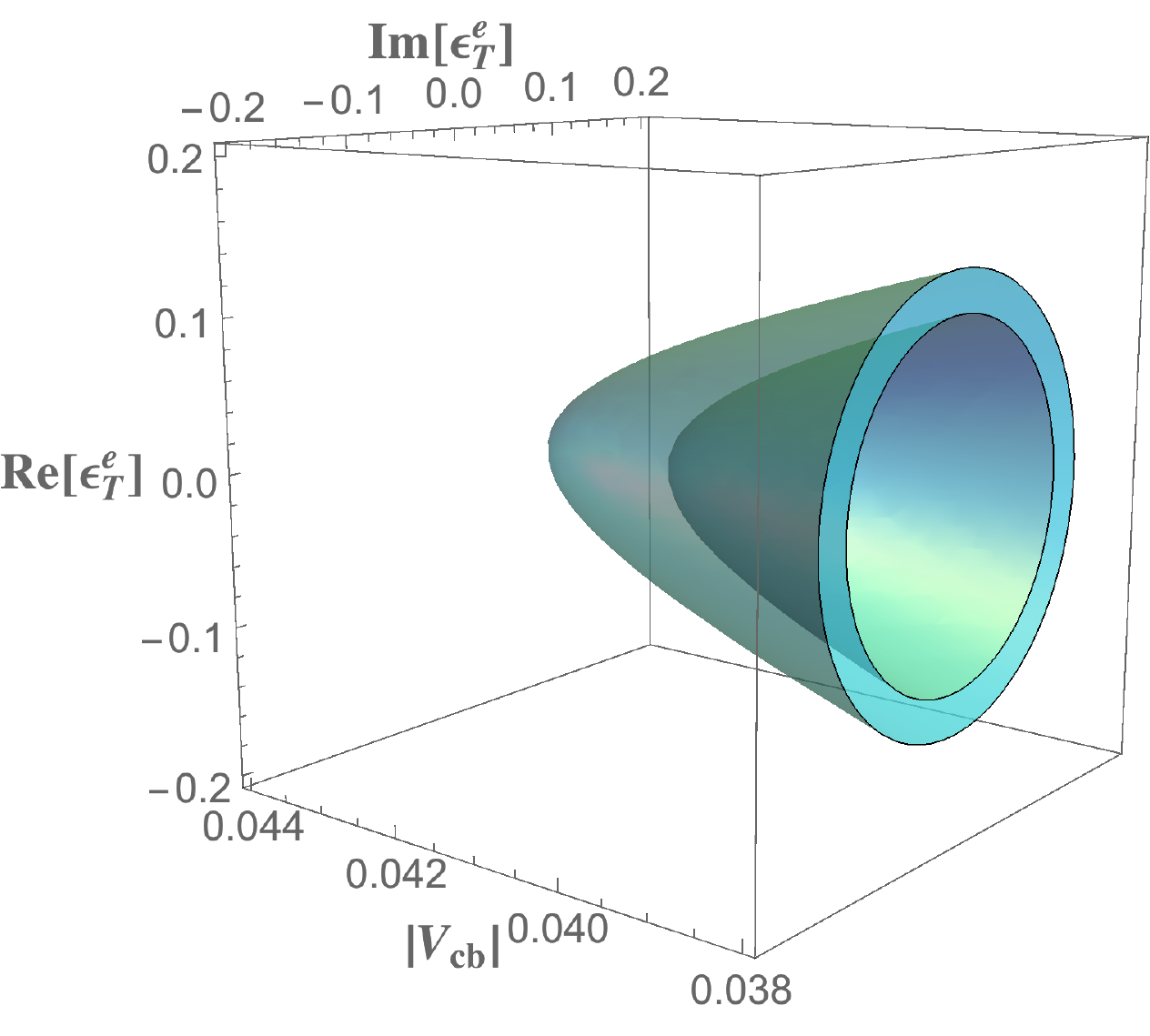}
\caption{Cutaway view of the parameter space ${\rm Re}(\epsilon_T^\ell)$, ${\rm Im}(\epsilon_T^\ell)$, and $\vcb$ allowing to obtain  ${\cal B}(B^+ \to X_c \ell^+ \nu_\ell)$ within $1 \sigma$ for muon (left) and electron mode (right).}\label{incl}
\end{figure}

Let us explain the reasoning  in \cite{Colangelo:2016ymy}.  The effective Hamiltonian is in (\ref{heff}), with the  assumption that not only $\epsilon_T^\tau$, but also $\epsilon_T^{(\mu,e)}$ can be non vanishing.
The observables are written as the sum of three terms:   the SM contribution,  the NP term  generated by the tensor operator  and  their interference.
For  the inclusive mode,   the spectrum in the dilepton invariant mass  ${\hat q}^2=q^2/m_b^2$ reads
\be
\frac{d\Gamma}{d\hat q^2}=C(q^2) \left[
\frac{d\tilde \Gamma}{d\hat q^2}|_{SM}+ |\epsilon_T|^2 \frac{d\tilde \Gamma}{d \hat q^2}|_{NP}+{\rm Re}(\epsilon_T)
\frac{d\tilde \Gamma}{d\hat q^2}|_{INT}\right]  ,  \label{incl-dgamma} 
  \ee
with
  $C(q^2)= \frac{G_F^2 |V_{cb}|^2 m_b^5}{96 \pi^3} \lambda^{1/2}\left(1-\frac{{\hat m}_\ell^2}{{\hat q}^2}\right)^2$,
   ${\hat m}_\ell=m_\ell/m_b$,  $\lambda=\lambda(1,\rho,{\hat q}^2)$ the triangular function and $\rho=m_c^2/m_b^2$.  
Using the Heavy Quark Expansion (HQE)
\cite{Chay:1990da}, each term   in (\ref{incl-dgamma})  can be written  as a series in powers of  the inverse heavy quark mass. In turn, each term in the series is the product of coefficient functions times matrix elements  of local operators with increasing dimension; at each order in $1/m_Q$, the coefficient functions 
can be  expanded in $\alpha_s$. The leading term reproduces the  free $Q$ decay width; the ${\cal O}(m_Q^{-1})$  term is absent. 
The SM result in the case of massive leptons can be found in \cite{Falk:1994gw}, the NP and INT terms  at leading order in $1/m_Q$ is in \cite{Biancofiore:2013ki} and at  ${\cal O}(m_Q^{-2})$ in \cite{Colangelo:2016ymy}.
Integrating (\ref{incl-dgamma}) the full width is obtained, so that the branching fraction can be compared to the measurement:
${\cal B}(B^+ \to X_c e^+ \nu_e) = (10.8 \pm 0.4) \times 10^{-2}$  \cite{Patrignani:2016xqp}.
Since no data are separately reported for muon, this result has to be used also in that case, while in the theoretical prediction  $m_e \neq m_\mu$ is kept.
The datum constrains $({\rm Re}(\epsilon_T^\ell),\,{\rm Im}(\epsilon_T^\ell),\,\vcb)$, as shown in 
Fig.~\ref{incl}, and  an upper bound on $\vcb$ can be established: 
at 1$\sigma$,    $\vcb \le 42.85 \times10^{-3}$ in the muon case  and $\vcb \le 42.73 \times10^{-3}$
in the electron case, in correspondence to the SM point  $\epsilon_T^\ell=0$. Nonvanishing values of $\epsilon_T^\ell$ produce the  regions in Fig.~\ref{incl}, which  continue  to smaller values of $\vcb$ if
  $|\epsilon_T^\ell |$ is increased.   $\vcb$ is bounded from below by the $B \to D^{(*)} \ell \bar \nu_\ell$ modes.
  %%%%%%%%%%%%%%%%%%%%%
%%%%%%%%%%%%%%%%%%%%%%%%%
The  description  of such modes requires the  $B\to D^{(*)}$ form factors that, in the HQ limit, can be related to  
the Isgur-Wise function $\xi(w)$ \cite{Isgur:1989ed}, with $w$  defined by the relation $q^2=m_B^2+m_{D^{(*)}}^2-2 m_B m_{D^{(*)}} w$.  $1/m_Q$  and ${\cal O}(\alpha_s)$ corrections can be included \cite{Neubert:1993mb,Caprini:1997mu}.  
For finite quark mass, several form factor determinations are available. 
In  \cite{Colangelo:2016ymy}  lattice QCD results  \cite{Lattice:2015rga,Bailey:2014tva} have been exploited. The other  form factors  have been derived using  HQ relations at  NLO, as described in \cite{Biancofiore:2013ki}. 
For  the decays  $B^- \to D^0 \ell^- \bar \nu_\ell$, the BaBar Collaboration has provided  separate measurements  for $\mu$ and $e$ modes \cite{Aubert:2008yv}:
$
{\cal B}(B^- \to D^0 \mu^- \bar \nu_\mu) = (2.25 \pm 0.04 \pm 0.17)\%$, $
{\cal B}(B^- \to D^0 e^- \bar \nu_e) = (2.38 \pm 0.04 \pm 0.15) \%$.
Comparison of  these data with theory  constrains $({\rm Re}(\epsilon_T^\ell),\,{\rm Im}(\epsilon_T^\ell),\,\vcb)$, while for
  $B^- \to D^{*0} \ell^- \bar \nu_\ell$,    the comparison is performed for the distribution $\displaystyle{\frac{d \Gamma}{dw}}$ close to the zero recoil point $w \to 1$ using the BaBar result  \cite{Aubert:2008yv}. 
  
   %%%%%%%%%%%%%%%%%%%%%
\begin{figure}[b!]
\vspace*{-0.2cm}
\includegraphics[width = 0.3\textwidth]{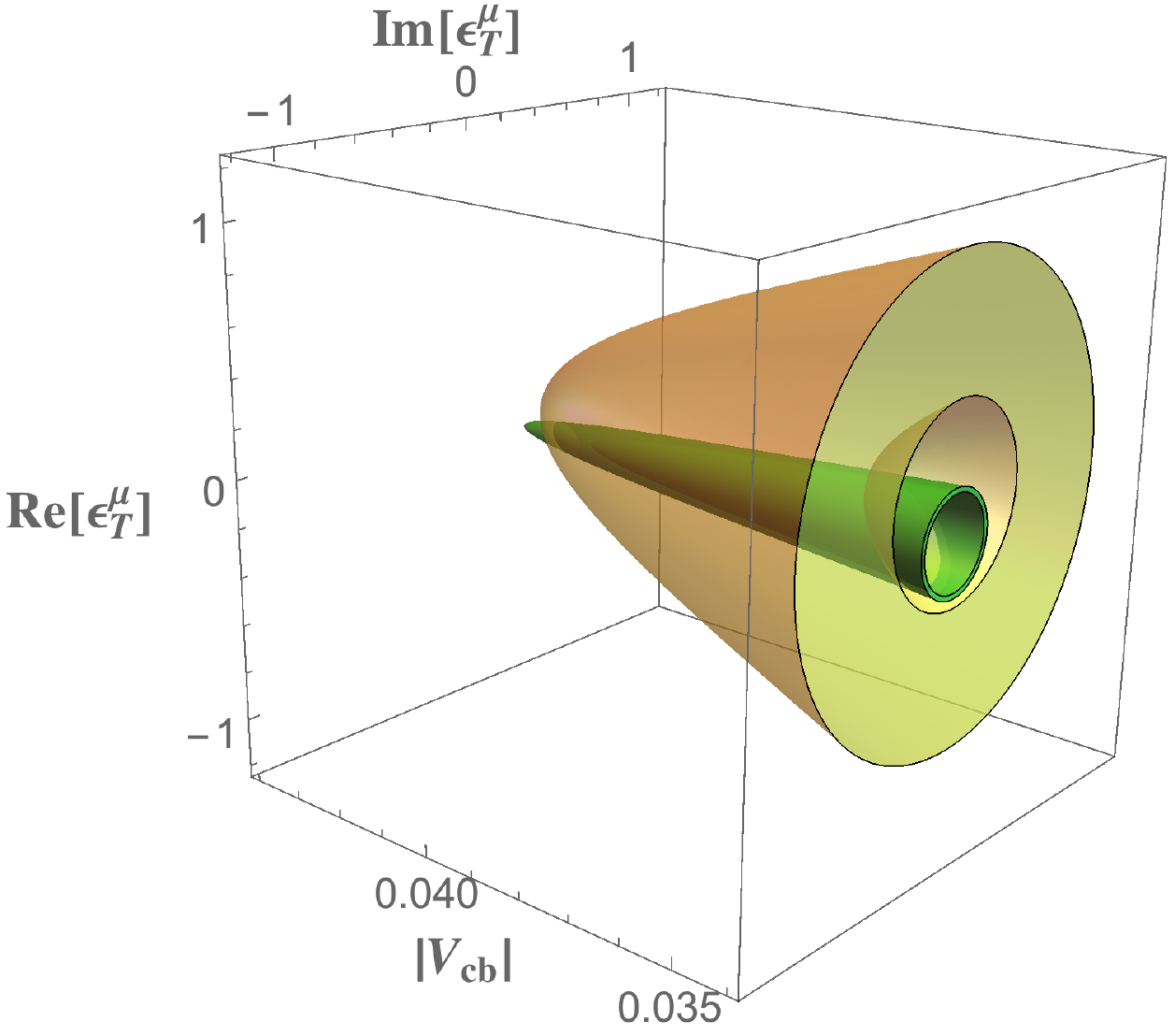} \hskip .5cm
\includegraphics[width = 0.3\textwidth]{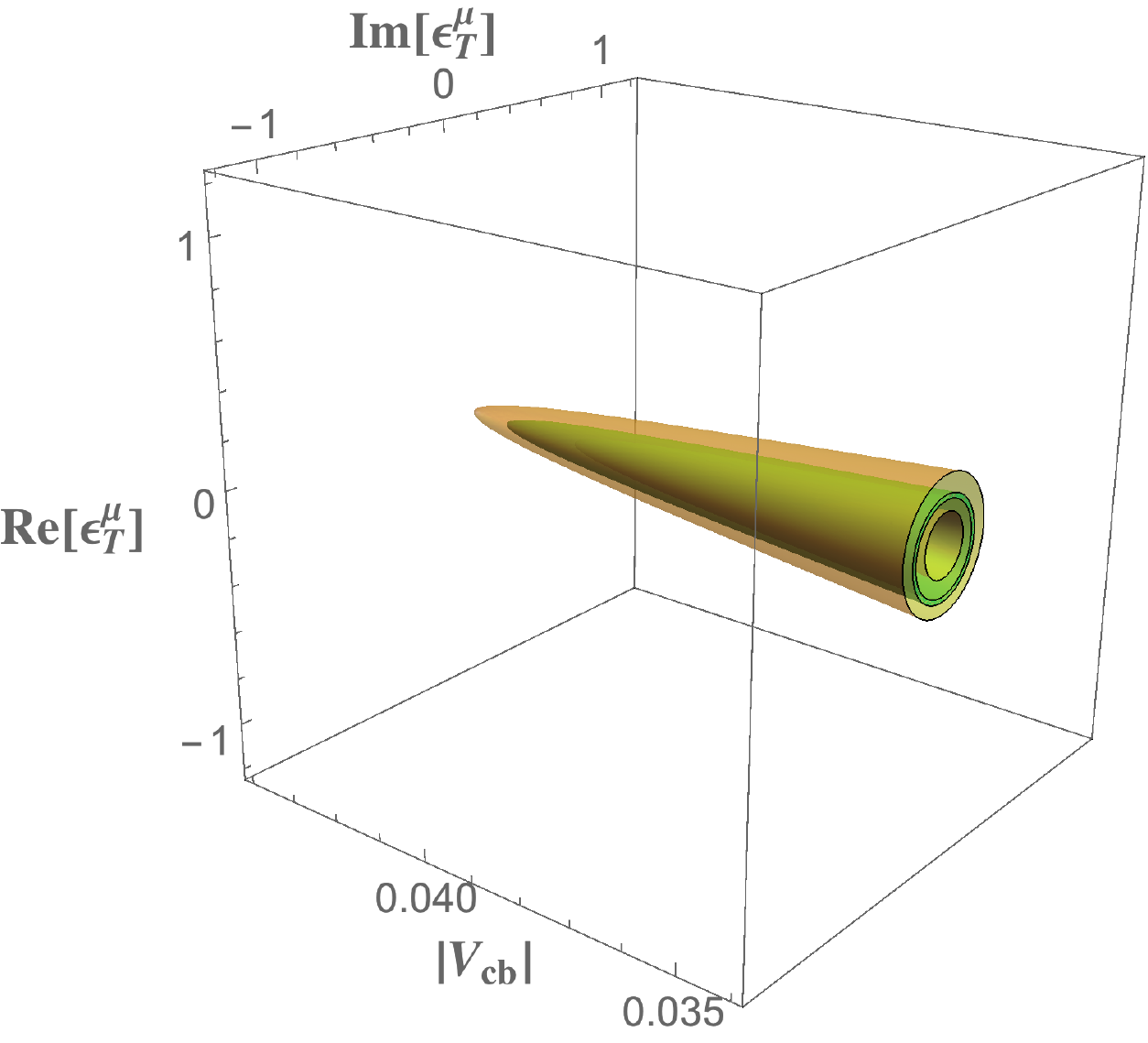} \hskip .5cm
\includegraphics[width = 0.3\textwidth]{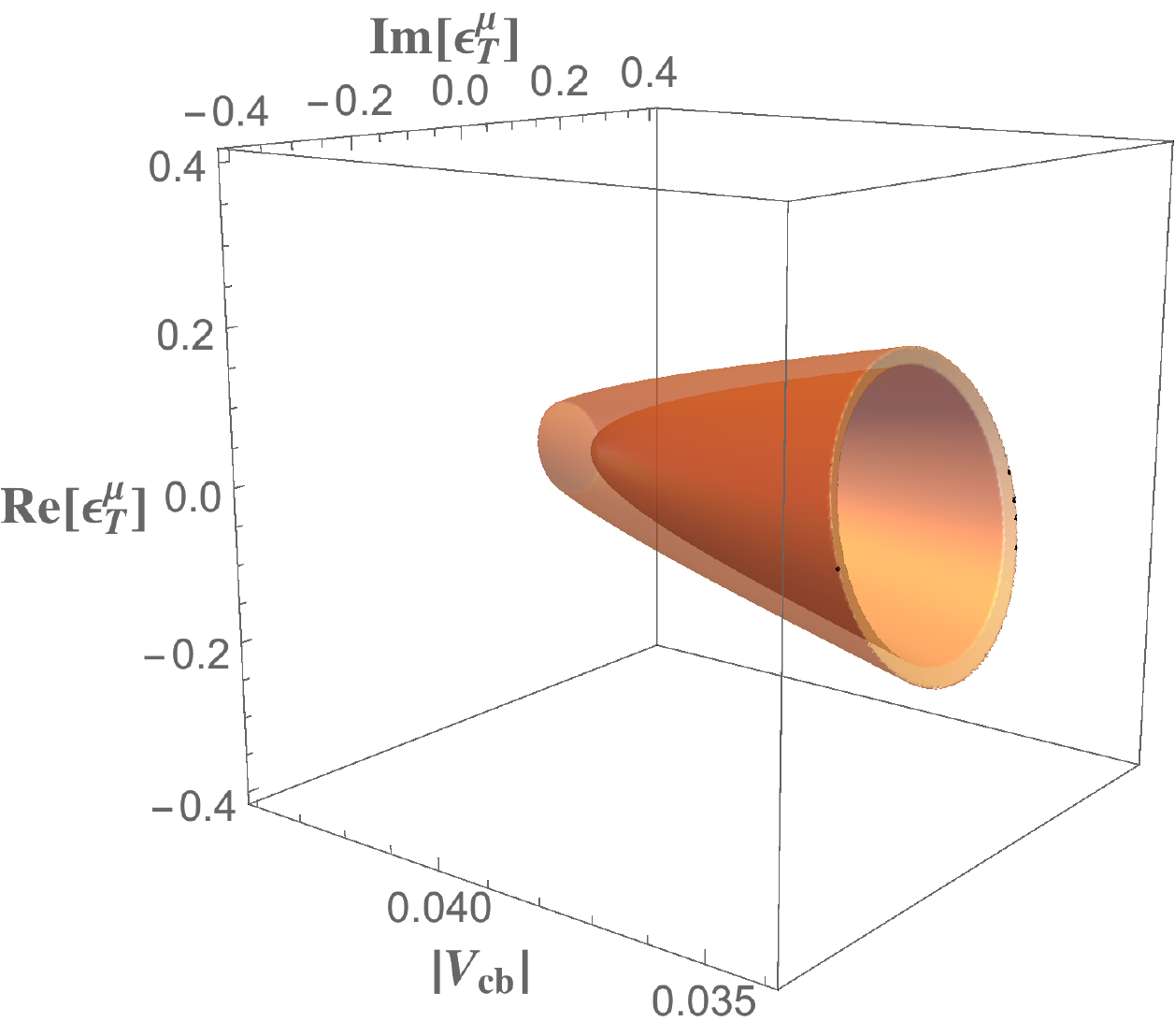}\\
\includegraphics[width = 0.3\textwidth]{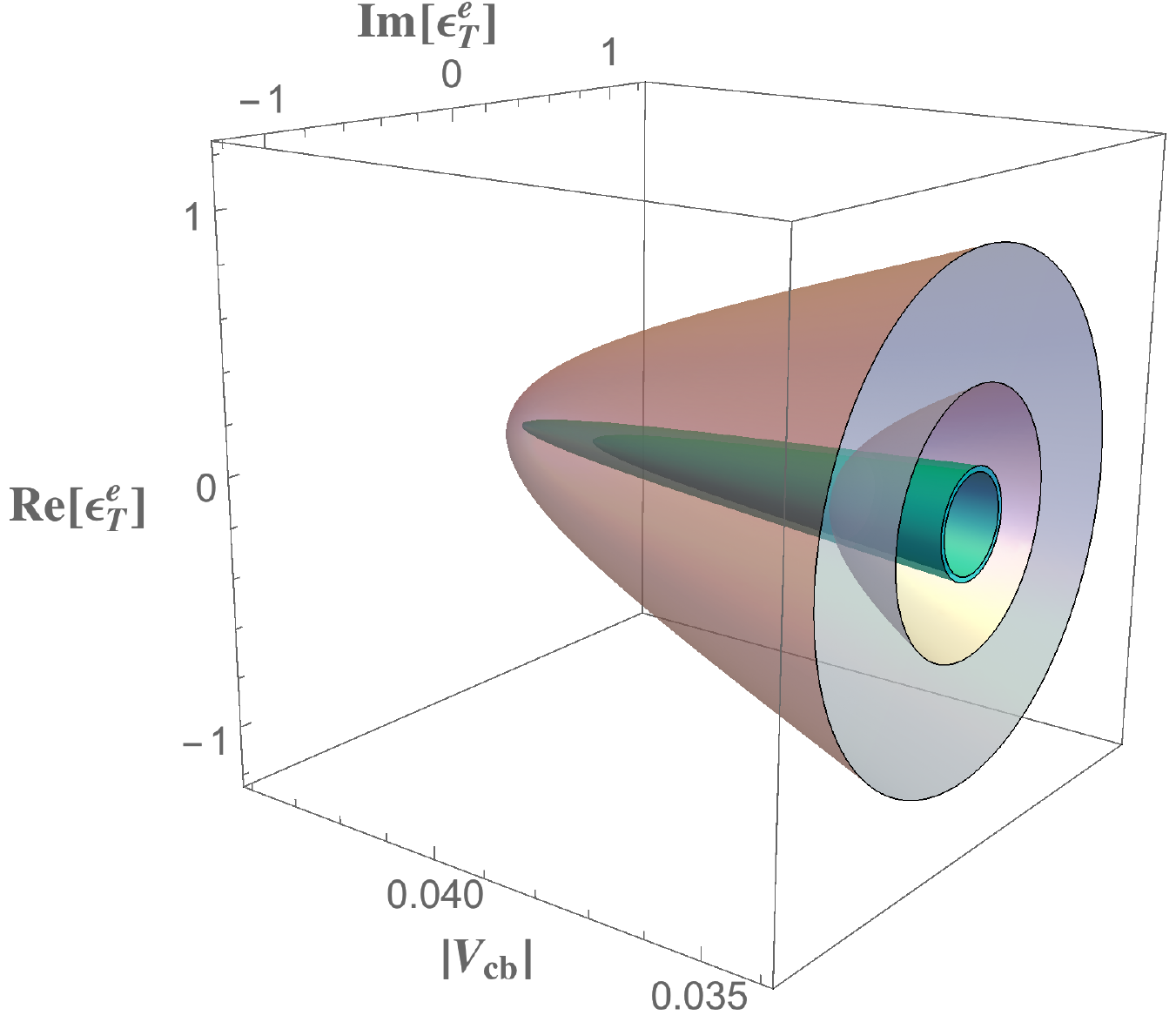}\hskip .5cm
\includegraphics[width = 0.3\textwidth]{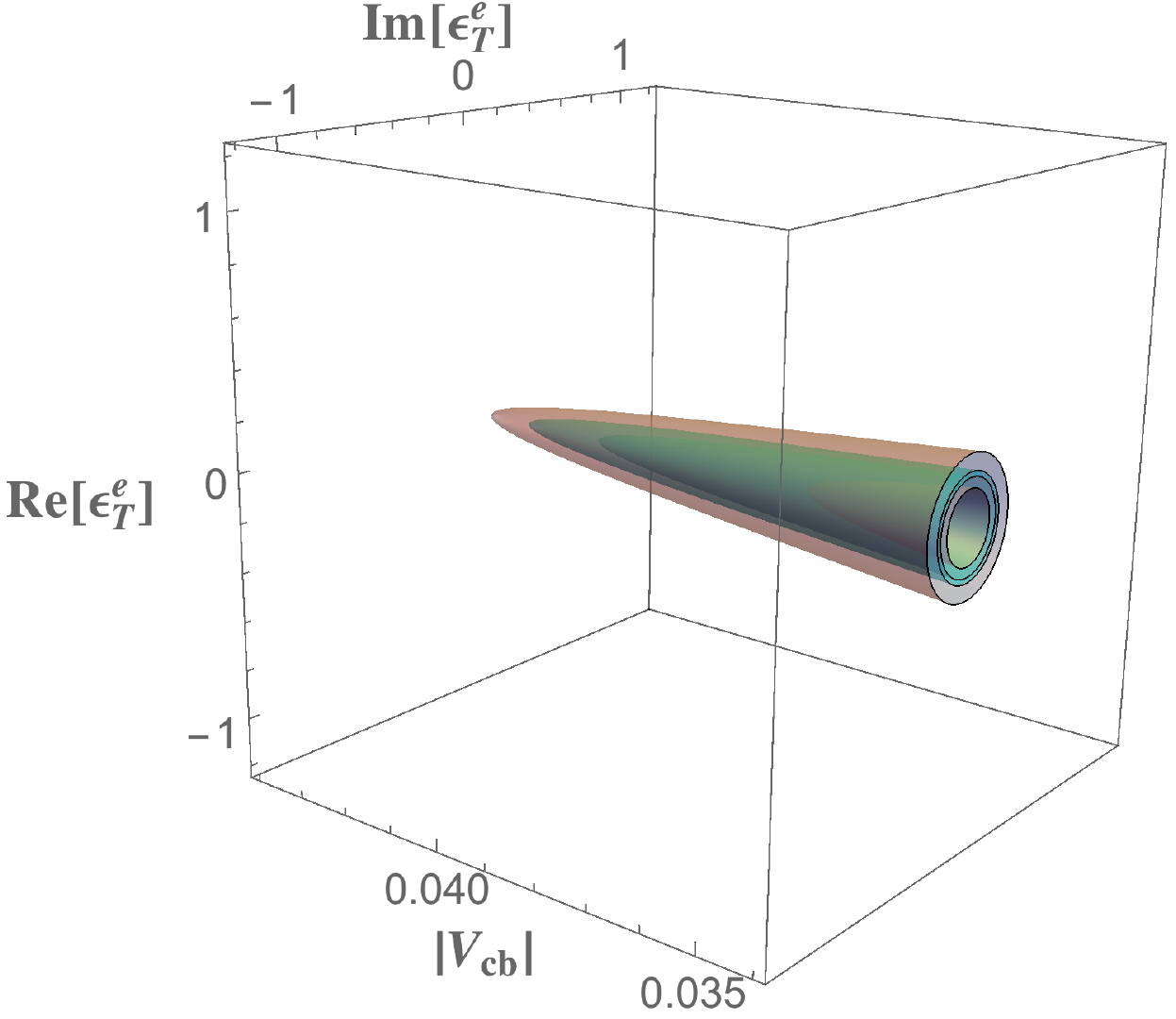}\hskip .5cm
\includegraphics[width = 0.3\textwidth]{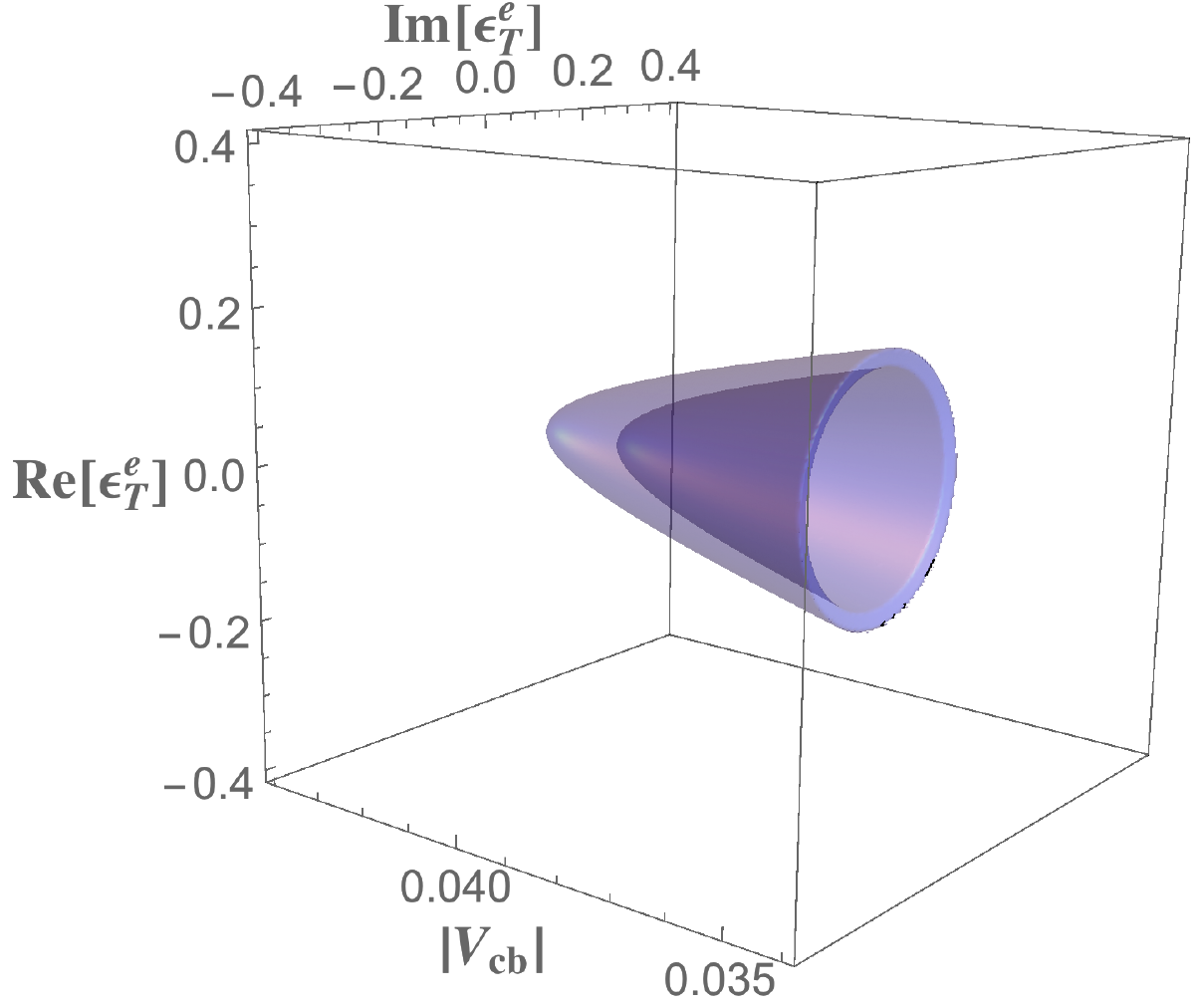}
\caption{Top panels: allowed regions in the $({\rm Re}(\epsilon_T^\mu),\,{\rm Im}(\epsilon_T^\mu),\,\vcb)$ parameter space, determined from  the inclusive mode (internal hollow region) together with  the  exclusive  $D$ mode (external hollow region, right plot), and from  the inclusive mode together with the decay to $D^*$   (external hollow region, middle plot).   
The intersection  of the three regions is shown in the right plot.
The bottom panels refer to the electron  case. }\label{fig:comb-mu}
\end{figure}
%%%%%%%%%%%%%%%%%%%%%%%%%
 %%%%%%%%%%%%%%%%%%%%%
\begin{figure}[b!]
\includegraphics[width = 0.59\textwidth]{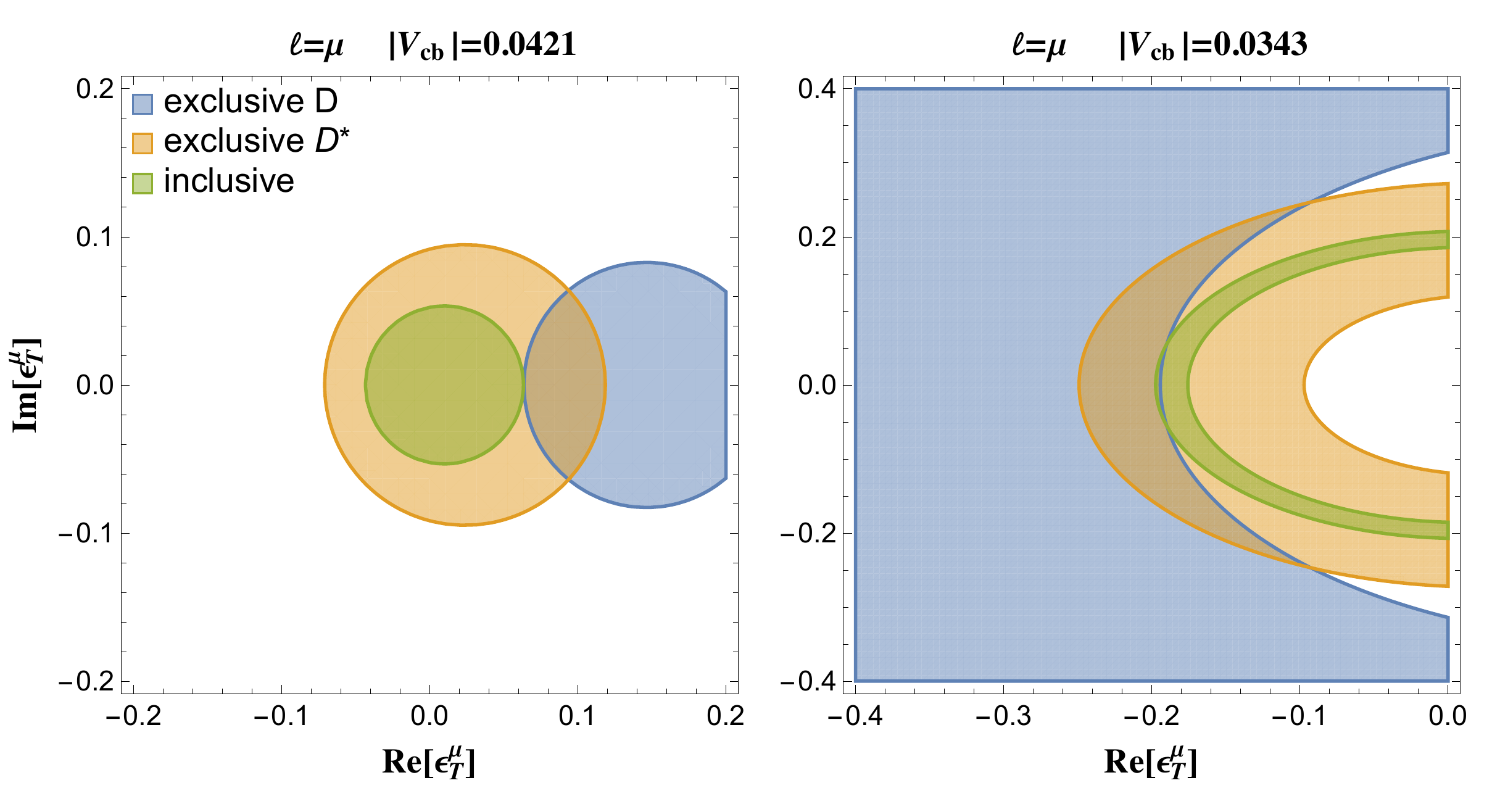}
\includegraphics[width = 0.39\textwidth]{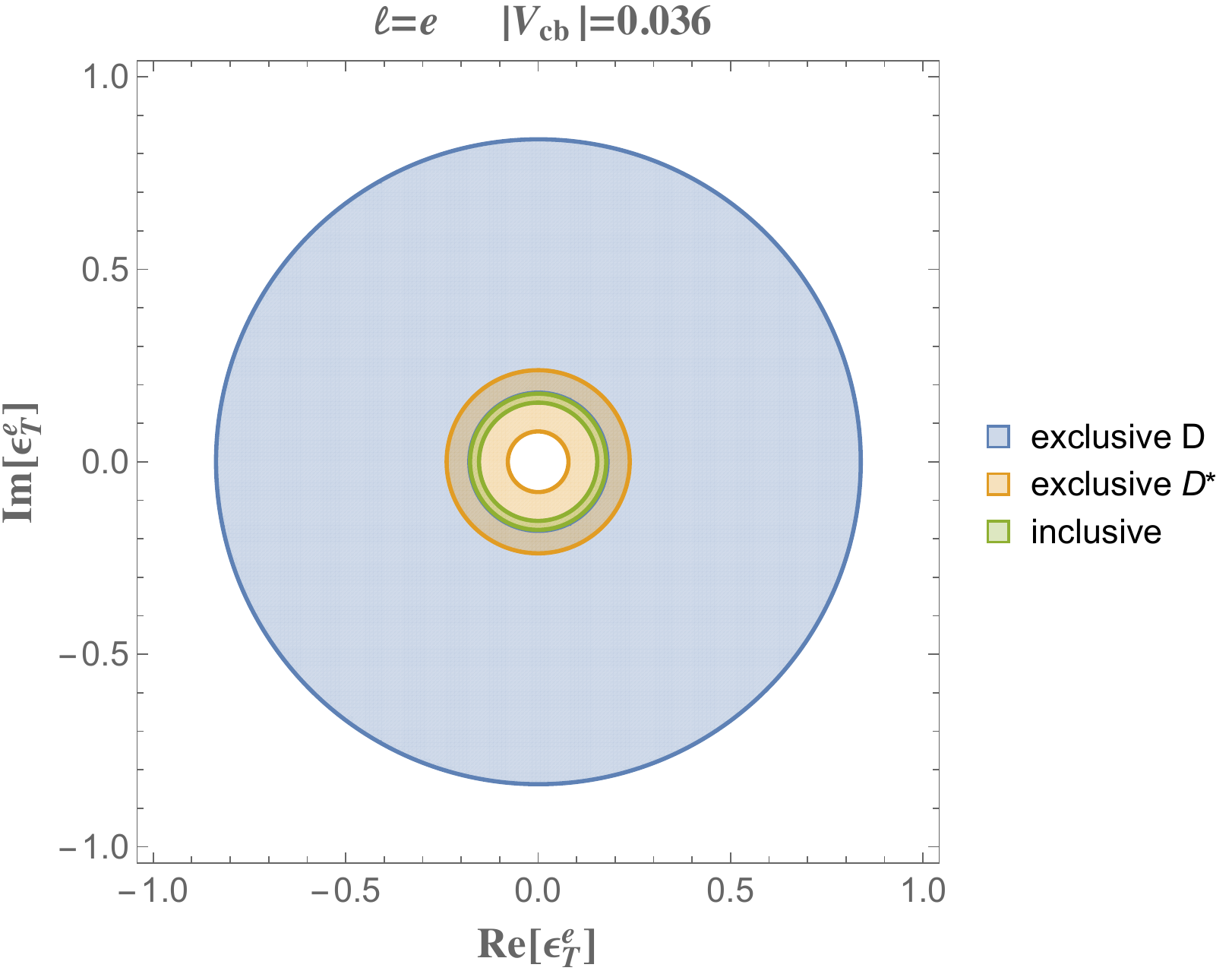}
\caption{Muon channel (left and middle panel): projections of the overlap  region in  
%Fig.~\ref{fig:comb-mu} 
Fig.~4 
on the $({\rm Re}(\epsilon_T^\mu),\,{\rm Im}(\epsilon_T^\mu))$ plane. 
The blue region corresponds to the constraint from the exclusive decay to $D$, the orange region  to the constraint from  $D^*$ and the green region to the constraint from the inclusive mode. The left (middle) plot corresponds to the largest (smallest) allowed  value of $\vcb$.  Electron channel (right plot): projections of the overlap region in 
%Fig.~\ref{fig:comb-mu} 
Fig.~4 
in the $({\rm Re}(\epsilon_T^e),\,{\rm Im}(\epsilon_T^e))$ plane.
  $\vcb$ is set to  the smallest allowed  value. Color coding as in the $\mu$ case.}\label{rpnew-mu}
\end{figure}
%%%%%%%%%%%%%%%%%%%%%%%%%
All the constraints  from   inclusive  and  exclusive   modes can be combined.
In the  muon channels,   the  upper (left and middle) panels  of 
Fig.~\ref{fig:comb-mu} 
show  the parameter regions selected by the exclusive constraints (larger yellow space)   superimposed to the region determined from the inclusive mode  (smaller inner space). The left plot refers to  ${\cal B}(B^- \to D^0 \mu^- \bar \nu_\mu)$,   the middle one to  $\displaystyle{\frac{d \Gamma}{dw}}(B^- \to D^{*0} \mu^- \bar \nu_\mu)$.
The  region where all  constraints are fulfilled is shown in the right plot. 
The bottom panels in the same figure refer to the electron mode.

In the case of $\mu$,  exclusive data  put a lower bound on $\vcb$ and slightly reduce the upper bound,  as   shown in Fig.~\ref{rpnew-mu} (left and middle plots). The figure  displays the projections in the $({\rm Re}(\epsilon_T^\mu),\,{\rm Im}(\epsilon_T^\mu))$ plane of the parameter space  
in correspondence to the extreme values of $\vcb$,  those for which the regions selected using the various constraints do not overlap.
The obtained  range is $\vcb \in [0.0343,\,0.0421]$. The range for $\epsilon_T^\mu$  depends on $\vcb$, with  the largest allowed value  $|\epsilon_T^\mu|\simeq 0.2$.
The lepton mass effect and   the interference term are at the origin of the fact that 
the symmetry axes of the two regions of parameters, resulting from the inclusive and the exclusive constraints, do not intersect  the $({\rm Re}(\epsilon_T^\mu),\,{\rm Im}(\epsilon_T^\mu))$ plane at the origin and  do not coincide. Hence, the  NP contribution differently affects the inclusive  and the two exclusive channels.

For  the electron mode,  changes in  the results  are due to the tiny value of the $e$ mass.   The exclusive constraints do not modify the upper bound $\vcb \le 0.04273$  put by  the inclusive analysis, and  a lower bound is  found, as shown in Fig.~\ref{rpnew-mu} (right). The  range  $\vcb \in [0.036,\,0.0427]$ and  the  bound $|\epsilon_T^e|\le 0.17$ are found.
The conclusion is that  it is possible to find sets of three parameters  fulfilling all the constraints both for $\mu$ and $e$.  For $\vcb$ this happens in the range  $\vcb \in [0.036,\,0.042]$. Although 
 the uncertainty on $\vcb$ is quite sizable, the analysis shows that it is possible to  reconcile its  inclusive and exclusive  determinations via a non SM  contribution.
 %%%%%%%%%%%%%%%%%%%%%%%%%
\begin{figure}[t!]
\vspace*{-0.4cm}
\hspace*{-0.65cm}
\includegraphics[width = 0.58\textwidth]{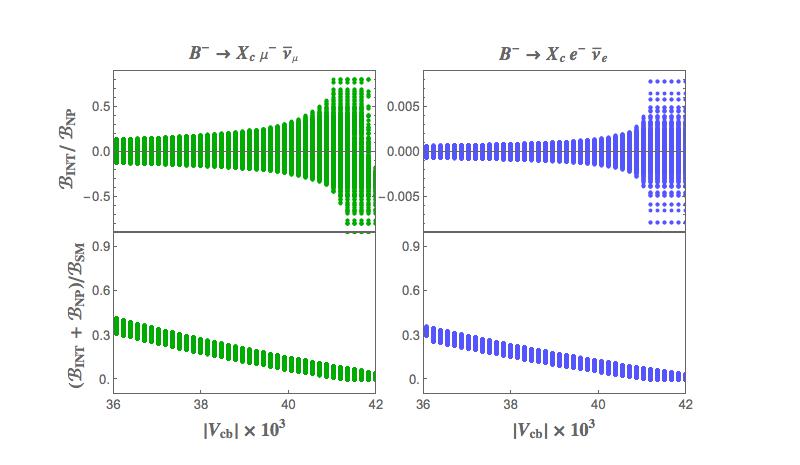}\hspace*{-0.7cm}
\includegraphics[width = 0.58\textwidth]{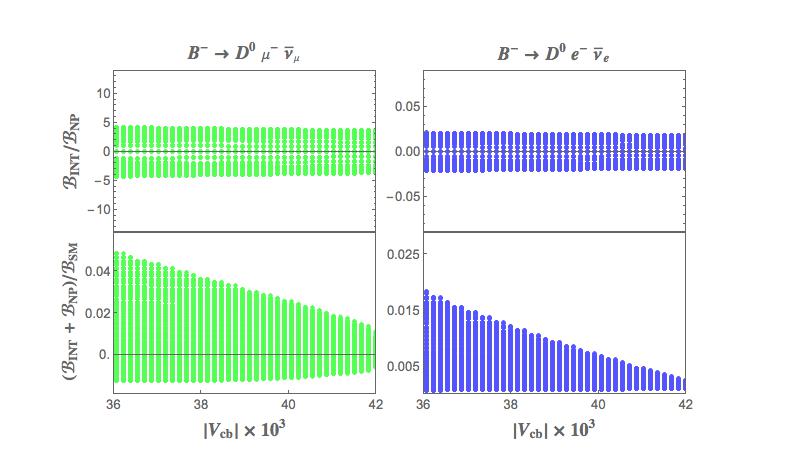}
\vspace*{-0.4cm}
\caption{Relative size of the NP contributions to ${\cal B}(B^- \to X_c \mu^- \bar \nu_\mu)$ (left panel, green) and ${\cal B}(B^- \to X_c e^- \bar \nu_e)$ (blue), and to ${\cal B}(B^- \to D^0 \mu^- \bar \nu_\mu)$ (right panel, green) and  ${\cal B}(B^- \to D^0 e^- \bar \nu_e)$ (blue).}\label{grid-incl-mu}
\end{figure}
%%%%%%%%%%%%%%%%%%%%%%%%%

The role of the NP contribution and of the interference  between  SM and NP can be assessed integrating  separately the three terms in  (\ref{incl-dgamma}). The  results, divided by the full decay width, are denoted by  ${\cal B}_{SM}$, ${\cal B}_{NP}$, ${\cal B}_{INT}$.  Fig.~\ref{grid-incl-mu} shows the ratios $\frac{{\cal B}_{INT}}{{\cal B}_{NP}}$ and $\frac{{\cal B}_{INT}+{\cal B}_{NP}}{{\cal B}_{SM}}$
 obtained varying $({\rm Re}(\epsilon_T^\ell),\,{\rm Im}(\epsilon_T^\ell),\,\vcb)$ in the allowed ranges. The left plots refer to the inclusive modes, the right ones to the exclusive decay to $D$. Green color refers to the muon, blue to  the electron. 
${\cal B}_{INT}$ can be sizable with respect to ${\cal B}_{NP}$ for the  muon and is  negligible for the electron. For both  $\mu$ and $e$,
 ${\cal B}_{INT}+{\cal B}_{NP}$
is negligible for large  $\vcb$ and sizable for small  $\vcb$.  The  tensor operator has a larger impact in the inclusive mode  than in  $D$ channel.  NP  affects in a different way  these decays: this is an example of how a new  term in the effective Hamiltonian  could  be at the origin of  the $\vcb$ anomaly.

%The ranges determined for    $\epsilon_T^{e}$ and  $\epsilon_T^{\mu}$ do not exclude that a NP contribution of the %kindconsidered here  still fulfills  $e-\mu$ universality.
%\vspace*{0.5 cm}

\section{Conclusions}
The anomalies emerged in $B$ decays seem to point to LFU violation. Several paths can be followed to confirm/discard this conclusion. The $R(D^{(*)})$ results suggest to search  similar effects  in $B_s$, $B_c$, $\Lambda_b$ semileptonic decays, as well as in $B$  to $D^{**}$ modes. The purely leptonic  $B_c \to \tau {\bar \nu}_\tau$ channel is a  testing ground for the scenario with a new tensor operator. The analyses would also contribute to shed light on the $\vcb$ puzzle: in this case,  separate rmeasurements for inclusive and exclusive semileptonic $B$  to $\mu$ and $e$ modes are called for \cite{Colangelo:2016ymy}.
To check the impact of form factor models, it is important to  investigate if there are analogous hints of LFU violation in  $b \to u$ transitions, also in connection with  the determination of $\vub$.

\vspace{0.2cm}
\noindent {\bf Acknowledgments.} I thank J.~Albrecht, D.~Becirevic and P.~Urquijo  for inviting me to present this talk at the EPS-HEP 2017 conference, and
P.~Biancofiore and P.~Colangelo for collaboration on the topics discussed here.
The work has been carried out within the INFN project QFT-HEP.

\end{document}